\documentclass[prd,reprint,amsmath,amssymb,longbibliography]{revtex4-1}
\pdfoutput=1
\usepackage{graphicx}
\usepackage[bookmarksopen,bookmarksnumbered]{hyperref}
\usepackage{color}
\usepackage[usenames,dvipsnames]{xcolor}
\usepackage{units}

\DeclareRobustCommand*\diff[2][]{%
   \mathop{
        \mathrm{d}^{#1}
     \mskip-0.2\thinmuskip
   #2}\nolimits
}

\let\oldleft\left
\def\xleft{\mathopen{}\oldleft}

\newcommand\ifrac[2]{{#1}/{#2}}

\providecommand{\T}[1]{}
\renewcommand{\T}[1]{\boldsymbol{#1}_{\text{T}}}
\newcommand{\Tj}[2]{\boldsymbol{#1}_{#2\,\text{T}}}
\newcommand{\Tsc}[1]{#1_{\text{T}}}

\begin{document}

%============================================

\title{Combining TMD factorization and collinear factorization}
\author{J. Collins}
\email{jcc8@psu.edu}
\affiliation{%
  Department of Physics, Penn State University, University Park PA 16802, USA
}
\author{L. Gamberg}
\email{lpg10@psu.edu}
\affiliation{Science Division, Penn State University Berks, Reading, Pennsylvania 19610, USA}
\author{A.~Prokudin}
\email{prokudin@jlab.org}
\affiliation{Science Division, Penn State University Berks, Reading, Pennsylvania 19610, USA}
\affiliation{Theory Center, Jefferson Lab, 12000 Jefferson Avenue, Newport News, VA 23606, USA}
\author{T.~C.~Rogers}
\email{trogers@odu.edu}
\affiliation{Department of Physics, Old Dominion University, Norfolk, VA 23529, USA}
\affiliation{Theory Center, Jefferson Lab, 12000 Jefferson Avenue, Newport News, VA 23606, USA}
\author{N.~Sato}
\email{nsato@jlab.org}
\affiliation{Theory Center, Jefferson Lab, 12000 Jefferson Avenue, Newport News, VA 23606, USA}
\author{B.~Wang}
\email{bowenw@mail.smu.edu}
\affiliation{Department of Physics, Old Dominion University, Norfolk, VA 23529, USA}
\affiliation{Theory Center, Jefferson Lab, 12000 Jefferson Avenue, Newport News, VA 23606, USA}

\date{December 31, 2016}

%============================================
\begin{abstract}
  We examine some of the complications involved when combining
  (matching) TMD factorization with collinear factorization to allow
  accurate predictions over the whole range of measured transverse
  momentum in a process like Drell-Yan.  Then we propose some improved
  methods for combining the two types of factorization.  (This talk is
  based on work reported in arXiv:1605.00671.)
\end{abstract}

\maketitle

%============================================
\section{Introduction}
\label{sec:intro}

This talk was based on \citet{Collins:2016hqq}, and provides a summary
of the results there.  More detailed references to earlier literature
can be found in that paper.

The issue addressed is the matching of transverse-momentum-dependent
(TMD) and collinear factorization for processes like the Drell-Yan
process, with both a hard scale $Q$ and a separate measured transverse
momentum $\Tsc{q}$.  TMD factorization applies when $\Tsc{q} \ll Q$, and
its accuracy degrades as $\Tsc{q}$ increases towards $Q$.  It involves
TMD parton densities (pdfs)
$f_{\text{parton}/\text{hadron}}(x,\Tsc{k})$, and in more general
process also TMD fragmentation functions.

In contrast, collinear factorization applies at large $\Tsc{q}$ (i.e.,
of order $Q$), and it also applies to the cross section integrated
over all $\Tsc{q}$ (and hence for the integral over $\Tsc{q}$ up to a
maximum of order $Q$).  Collinear factorization involves ``collinear
pdfs'' $f_{\text{parton}/\text{hadron}}(x)$.  Its accuracy degrades as
$\Tsc{q}$ decreases, and collinear factorization by itself provides
unphysically singular cross sections as $\Tsc{q}\to0$.  But the nature
of the degradation is constrained by the fact the collinear
factorization also applies to the cross section integrated over
$\Tsc{q}$.

To get full accuracy over all $\Tsc{q}$ one may combine both methods
suitably.  Collins, Soper, and Sterman (CSS) \cite{Collins:1981va,
  Collins:1984kg} implemented this as a kind of simple matched
asymptotic expansion.  But it has become increasingly clear---e.g.,
\cite{Boglione:2014oea, Boglione:2014qha}---that improved matching
methods are needed to get adequate performance in practice, especially
at the relatively low $Q$ used in many experiments on semi-inclusive
deep-inelastic scattering (SIDIS).  These are of particular relevance
to this conference, since these experiments measure important
transverse spin asymmetries analyzed with the aid of TMD
factorization, as can be seen from other contributions at the
conference.

We will summarize the issues and then our proposed improvements in the
matching methods.

%============================================
\section{Key approximations to get TMD and collinear factorization}

\begin{figure}
  \centering
   \includegraphics[scale=0.45]{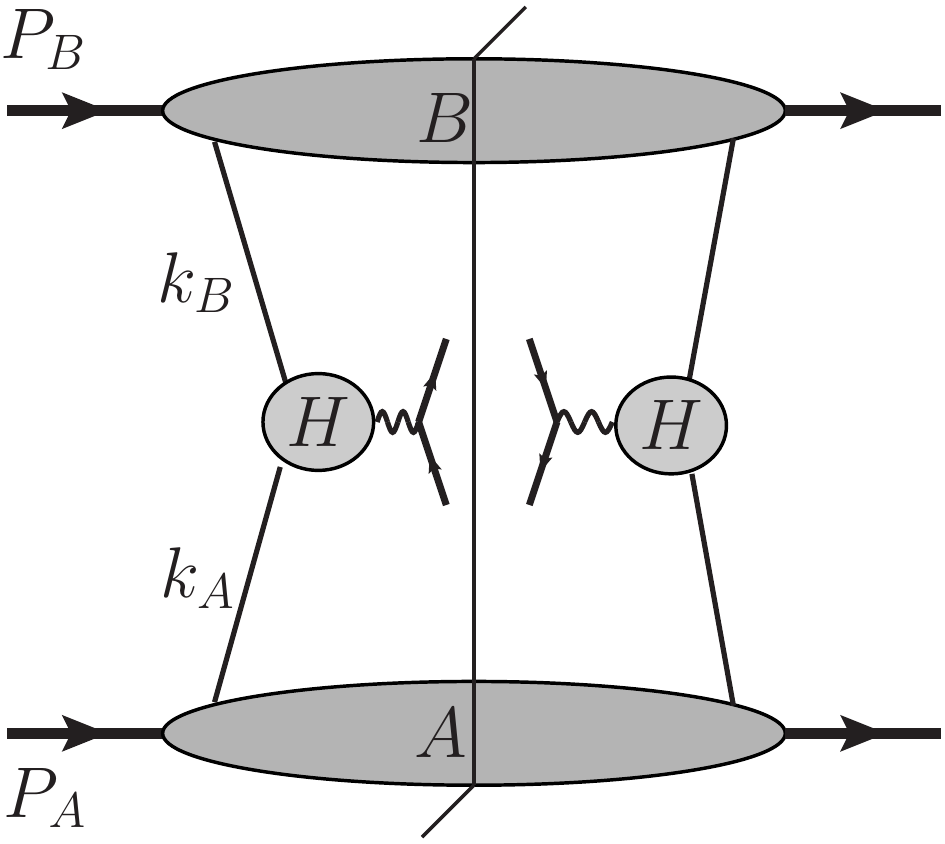}
\caption{Parton-model structure for Drell-Yan process.}
\label{fig:parton.model.graph}
\end{figure}

The essential measure of the quality of the applicability of a
matching method is an evaluation of its accuracy.  This is determined
by the accuracy of the approximations used in deriving factorization
from the exact cross section.  The approximations can be understood
from an examination of the derivations, as in
\cite{Collins:2011qcdbook}.  A simple example is given by the
Feynman-graphical structure of the basic parton-model form, Fig.\
\ref{fig:parton.model.graph}, where the Drell-Yan pair is created by
quark-antiquark annihilation, with the quark and antiquark arising
from structures that are collinearly moving with respect to the
incoming hadrons.

We use light-front coordinates, where the parton momenta are
\begin{equation}
  k_A=(x_AP_A^+, k_A^-, \Tj{k}{A}),
  \quad k_B=(k_B^+,x_BP_B^-, \Tj{k}{B}). 
\end{equation}
The incoming hadrons $p_A$ and $p_B$ have large 3-momenta
in the $+z$ and $-z$ directions, with no transverse momentum.

%----------------------------
\subsection{TMD factorization}

To get the corresponding contribution to a TMD-factorized form,
approximations are made: (a) In the hard-scattering subgraph, $H$, the
exact parton momenta $k_A$ and $k_B$ are replaced by on-shell values.
(b) But in the kinematics, parton transverse momentum is retained, so
that the virtual photon momentum $q$ is
$(x_AP_A^+,x_BP_B^-,\Tj{k}{A}+\Tj{k}{B})$.  Thus, after the
approximations, the dependence on the small components $k_A^-$ and
$k_B^+$ is confined to the subgraphs $A$ and $B$, respectively, while
the transverse momentum of the Drell-Yan pair arises from the quark
transverse momenta.

We can then integrate over $k_A^-$ within $A$ and $k_B^+$ within $B$,
to obtain the natural contributions to the usually defined TMD pdfs.
The approximations are valid because $k_A^- = O(\Tsc{q}^2/q^+)$, etc.
The approximations become bad when $\Tsc{q}$ increases to roughly
order $Q$.

\begin{figure}
  \centering
  \includegraphics[scale=0.45]{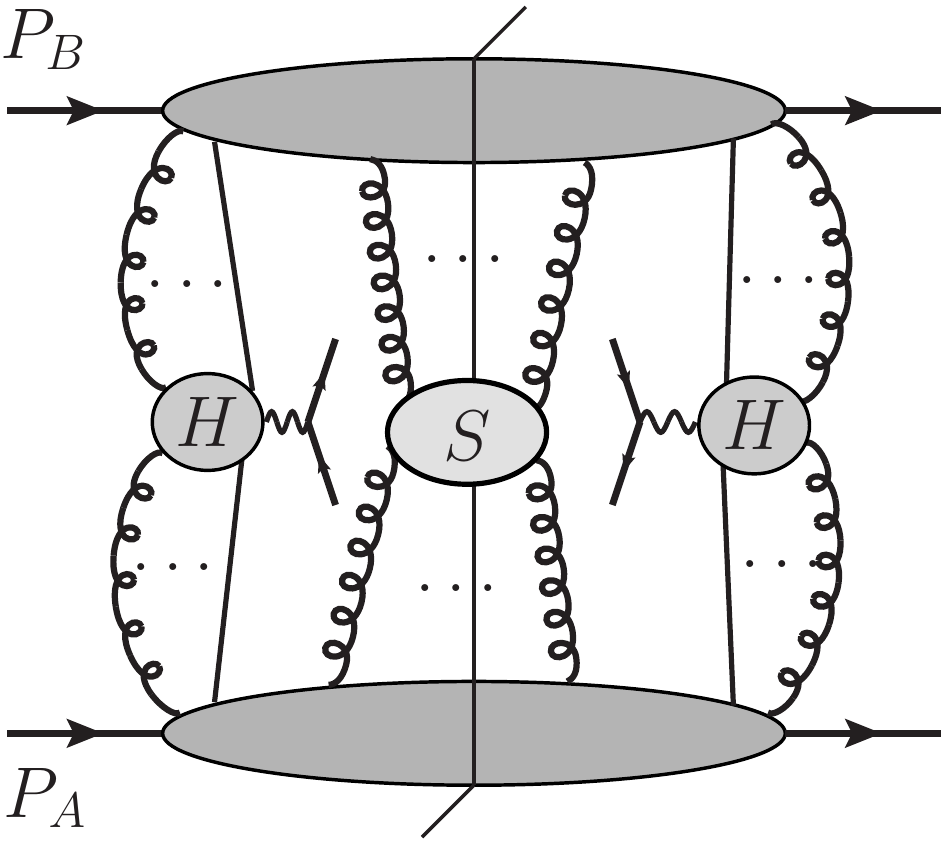}
  \caption{Structures for Drell-Yan to leading power at low
    $\Tsc{q}$.}
  \label{fig:DY.leading}
\end{figure}

Other graphical structures giving leading power contributions, Fig.\
\ref{fig:DY.leading} are treated similarly, with the application of
Ward identities and unitarity-style cancellations to get the
TMD-factorized form.

%----------------------------
\subsection{Collinear factorization for large \texorpdfstring{
    $\Tsc{q}$}{qT}, and for integral over \texorpdfstring{
    $\Tsc{q}$}{qT}}

\begin{figure}
  \centering
  \includegraphics[scale=0.45]{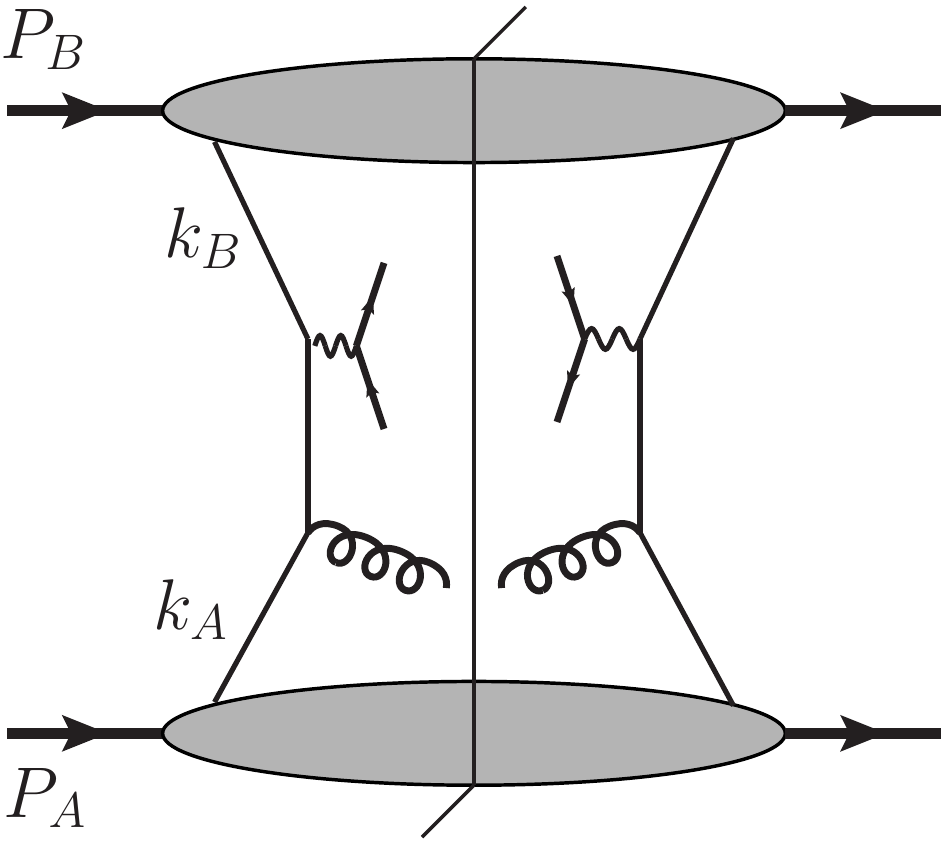}
  \caption{Example of structure giving Drell-Yan at large $\Tsc{q}$.}
  \label{fig:DY.large.qT}
\end{figure}

Large $\Tsc{q}$ is dominantly generated from hard scatterings where
extra partons are emitted, exemplified in Fig.\ \ref{fig:DY.large.qT}.
The appropriate leading-power approximation for the hard scattering
now neglects the transverse momenta of the incoming partons $k_A$ and
$k_B$, as well as their virtuality.  Each collinear pdf $f(x)$ is
therefore defined with an integral over transverse momentum, and
therefore depends kinematically only on the longitudinal momentum
fraction $x$ of the parton.

The collinear approximation involves neglecting small transverse
momenta of the incoming partons in comparison with $\Tsc{q}$, as well
as neglecting their virtualities.  Therefore the approximation
completely breaks down once $\Tsc{q}$ is of order a typical transverse
momentum for the partons entering the hard scattering.  A symptom of
the breakdown is the well-known strong singularity at $\Tsc{q}=0$ of
fixed-order calculations of the Drell-Yan cross section.

Next we turn to the cross section integrated over $\Tsc{q}$.  Here one
must include all the contributions at low $\Tsc{q}$.  But now the
collinear approximation remains valid, unlike the TMD case.  In graphs
like Fig.\ \ref{fig:parton.model.graph}, the collinear approximation
ignores the partonic $\Tsc{k}$ in the hard scattering, which shifts
the virtual photon's transverse momentum to zero from its true value.  But
since this is just a shift, it leaves the integral over $\Tsc{q}$
unchanged, to leading power in the large scale $Q$.  Thus although
collinear factorization is incorrect at low $\Tsc{q}$ for the
distribution in $\Tsc{q}$, it is nevertheless valid for the integral
over $\Tsc{q}$.

%----------------------------
\subsection{Error sizes}

\begin{figure}
  \centering
  \includegraphics[scale=0.45]{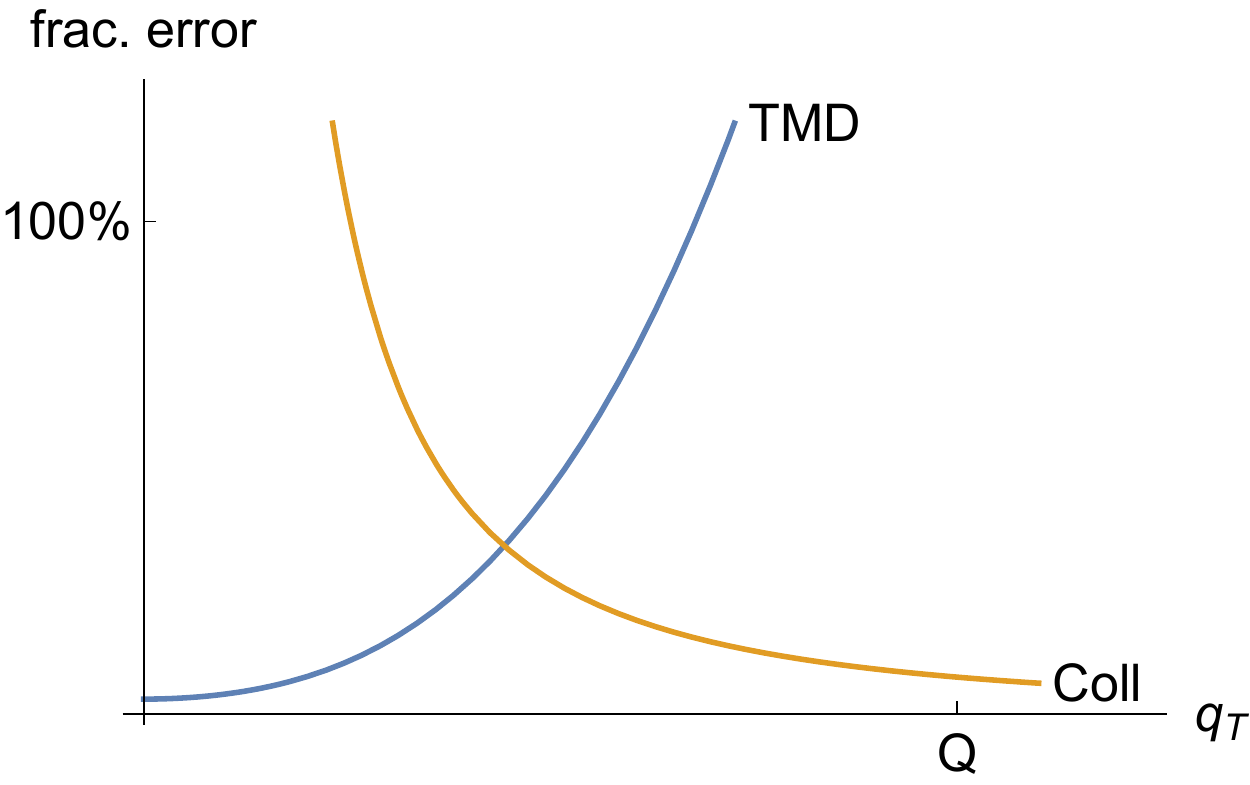}
  \caption{Qualitative behavior of fractional errors in TMD and
    collinear factorization as a function of $\Tsc{q}$, at fairly low $Q$.}
  \label{fig:error}
\end{figure}

The qualitative behavior of the fractional errors in TMD and collinear
factorization is shown in Fig.\ \ref{fig:error}.  TMD factorization is
accurate at low $\Tsc{q}$ up to relative errors suppressed by a power
of $1/Q$, but it is totally inaccurate at $\Tsc{q}$ of order $Q$.
Collinear factorization for the $\Tsc{q}$ distribution has the
opposite behavior.

%============================================
\section{CSS's \texorpdfstring{$W+Y$}{W+Y} method to combine TMD and
  collinear factorization}
\label{sec.error}

CSS implemented the combination of TMD and collinear factorization by
\begin{equation}
\label{eq:W+Y.CSS}
  \frac{ \diff{\sigma} }{ \diff[4]{q} } = W + Y + \mbox{error},
\end{equation}
where $W$ is the TMD factorized form
\begin{multline}
  W = \sigma_0 H(Q/\mu) \int \diff[2]{\T{b}} e^{i\T{q}\cdot \T{b} } 
\\
      \tilde{f}(x_A,\T{b}; \mu,Q) \tilde{f}(x_B,\T{b}; \mu,Q),
\end{multline}
and $Y$ is a collinear correction term
\begin{equation}
  Y = \mbox{collinear approx.\ to} 
      \left( \frac{ \diff{\sigma} }{ \diff[4]{q} } - W \right).
\end{equation}
In $W$, the convolution over the two TMD pdfs is rewritten as a Fourier
transform over a transverse position variable $\T{b}$.

The errors in $W$ and $Y$ caused by the approximations in the
derivations can be estimated as:
\begin{equation}
  W = \frac{ \diff{\sigma} }{ \diff[4]{q} } 
      \left\{ 1 + O\xleft[ \left(\frac{\Lambda}{Q}\right)^a \right]
               + O\xleft[ \left(\frac{\Tsc{q}}{Q}\right)^a \right]
      \right\}
\end{equation}
for $W$ when $\Tsc{q} \lesssim Q$,
\begin{equation}
  Y = \left( \frac{ \diff{\sigma} }{ \diff[4]{q} }  - W \right)
      \left\{ 1 + O\xleft[ \left(\frac{\Lambda}{\Tsc{q}}\right)^a \right]
               + O\xleft[ \left(\frac{\Lambda}{Q}\right)^a \right]
      \right\}
\end{equation}
for $Y$ when $\Lambda \lesssim \Tsc{q} \lesssim Q$.  Here $a$ is some fixed positive
number determined by QCD and the power-law errors in the
approximations used in deriving factorization. To these errors are to
be added truncation errors of perturbative calculations.  Hence the
error in Eq.\ \eqref{eq:W+Y.CSS} is estimated to be a uniform power of
$1/Q$ by
\begin{equation}
  \mbox{error}
  = \frac{ \diff{\sigma} }{ \diff[4]{q} } - W - Y
  = \frac{ \diff{\sigma} }{ \diff[4]{q} }
    \times
    O\xleft[ \left(\frac{\Lambda}{Q}\right)^a \right]
\end{equation}
over the range $\Lambda \lesssim \Tsc{q} \lesssim Q$.  As one increases $\Tsc{q}$ from
order $\Lambda$, the deviation of $W$ from the cross section increases.  But
a collinear approximation can be applied to this deviation.  As
$\Tsc{q}$ increases the deviation becomes larger, while its collinear
approximation becomes better, in proportion.

It is very important that the stated error estimate for TMD
factorization applies only when $\Tsc{q} \lesssim Q$, and can completely fail
for higher $\Tsc{q}$.  Similarly the stated error for collinear
factorization applies only when $\Tsc{q} \gtrsim \Lambda$, and can completely fail
at lower $\Tsc{q}$.

%============================================
\section{What is problematic with the original
  \texorpdfstring{$W+Y$}{W+Y} formulation?}

\begin{figure}
  \centering
  \includegraphics
         [clip=true,viewport=150 55 580 462,scale=0.40]%
         {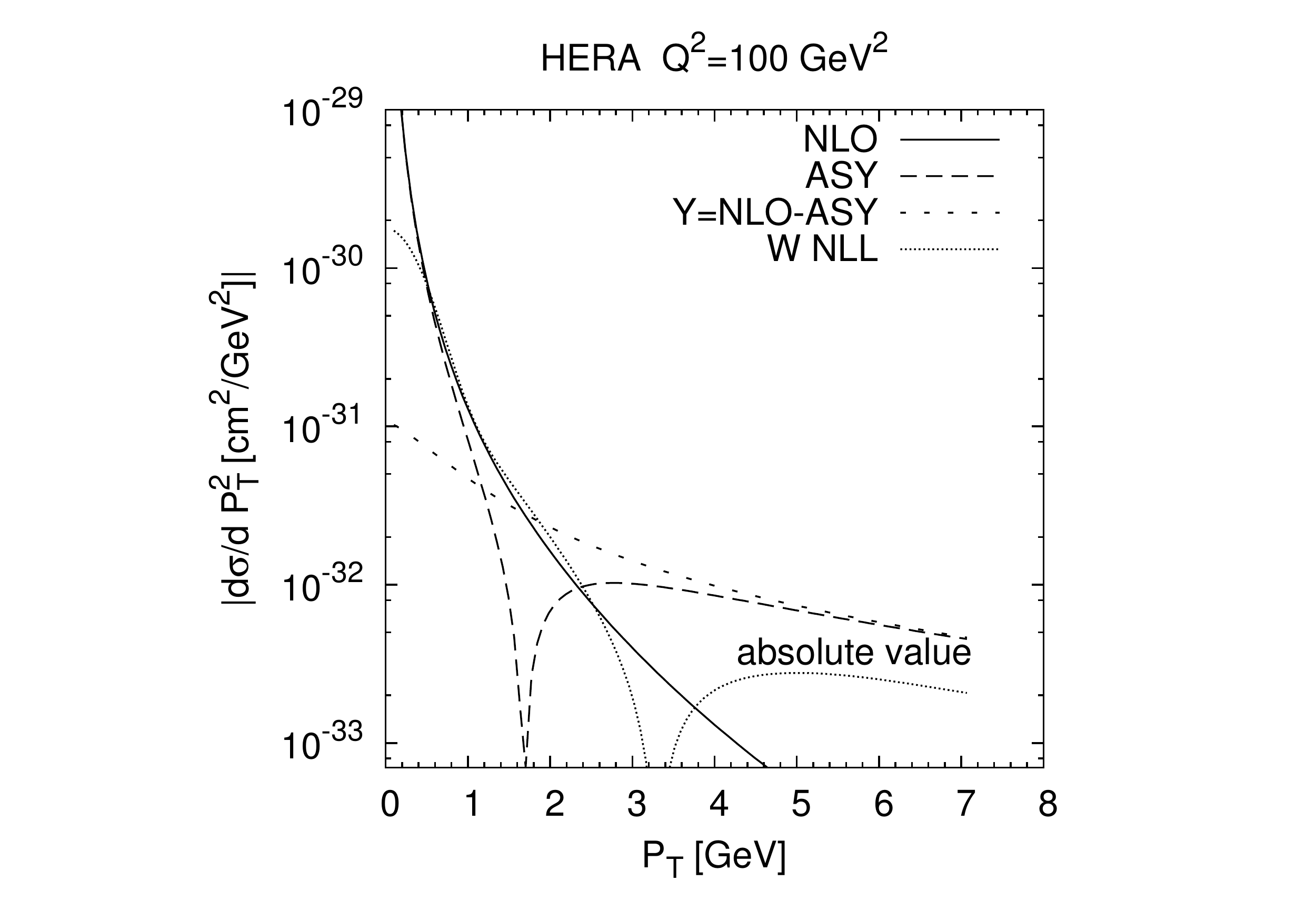}
         \caption{From Fig.\ 3 of Ref.\ \cite{Boglione:2014qha}, for
           SIDIS at HERA at $Q=\unit[10]{GeV}$, with $\Tsc{P}$ being
           the transverse momentum of the detected hadron in the
           $\gamma^*N$ frame. Some values (NLO and $W$) become negative at
           when $\Tsc{P}$ is increased enough; for these, the absolute
           values are plotted.}
  \label{fig:implementation}
\end{figure}

\citet{Boglione:2014oea, Boglione:2014qha} have found particular
difficulties with implementing CSS's $W+Y$ method in SIDIS at
moderately low energies.  These are illustrated in Fig.\
\ref{fig:implementation}, for SIDIS at $Q=\unit[10]{GeV}$ at HERA.
The solid line is the NLO calculation from collinear factorization. It
shows the decrease of the cross section with hadron transverse
$\Tsc{P}$ that QCD predicts at the larger values of $\Tsc{P}$.  But
the NLO curve diverges as $\Tsc{P}\to0$, where fixed-order calculations
in collinear factorization are totally inaccurate.

The dotted curve shows the $W$ term, i.e., the result of TMD
factorization, with fits to data determining the non-perturbative part
of the TMD functions and their evolution.  At small enough $\Tsc{P}$,
it should be a good approximation by itself for the cross section.

The asymptotic low $\Tsc{q}$ part of the NLO collinear-factorization
term is given by the dashed curve labeled ``ASY''.  This reproduces the
NLO calculation at low $\Tsc{P}$, but deviates from it substantially
as $\Tsc{P}$ increases.  The deviation becomes substantial at quite a
large factor below $Q$, which shows that there is a substantial
numerical degradation of the simplest error estimates that were given
in Sec.\ \ref{sec.error}.  The ASY term goes through zero at around
$\unit[1.7]{GeV}$ and then becomes negative.

Since the basic low $\Tsc{P}$ asymptote has a logarithm:
$\alpha_s\ln(Q/\Tsc{P})/\Tsc{P}^2$, the negative values are expected, and
are in a region where the asymptotic calculation is inapplicable.  The
bothersome issue is that the inapplicability happens at what appears
to be a surprisingly low $\Tsc{P}$ compared with $Q$.  Similarly $W$
goes through zero; this is also expected.  The principle of the CSS
method is that there should be a region of intermediate $\Tsc{P}$ or
$\Tsc{q}$ where neither TMD nor collinear factorization is completely
degraded in accuracy, at least for high $Q$.  In this case, the large
$\Tsc{P}$ part of the TMD factorization and the low $\Tsc{P}$ part of
collinear factorization should approximately match.  But this does not
happen in Fig.\ \ref{fig:implementation}.  Furthermore the zeros in
$W$ and ASY happen at quite different $\Tsc{P}$.  

The $Y$ term is the difference between the fixed-order collinear
calculation, in this case NLO, and its small-$\Tsc{P}$ asymptotic ASY.
CSS intended this to correct TMD factorization to collinear
factorization at large $\Tsc{q}$ or $\Tsc{P}$.  It is calculated to be
small when $\Tsc{P}$ is small, as it should be.  But it quickly
becomes much larger than the presumably approximately correct NLO
estimate for the cross section, which rather invalidates its use as a
correction.

These plots suggest some ideas for improving the $W+Y$ method.

Perhaps the most important practical problem is that the TMD term,
$W$, goes negative at large $\Tsc{P}$.  This indicates that at large
$\Tsc{P}$, $W+Y$ is the difference between substantially larger terms,
and therefore shows a strong magnification of the relative effects of
truncation errors in the predicted perturbative parts of the cross
section.

Associated with this is that the integral over $W$ is exactly zero.
This is because in $\Tsc{b}$-space, the evolution equations show that
the integrand is zero: $\tilde{W}(\Tsc{b})=0$.

From these properties arises a severe problem in getting the
integral over $\Tsc{q}$ of the $W+Y$ formula in Eq.\
\eqref{eq:W+Y.CSS} to agree with the collinear factorization results
\begin{itemize}
\item On the left-hand side, the integral $\int \diff[2]{\T{q}} \ifrac{
    \diff{\sigma} }{ \diff[4]{q}}$ is given by collinear factorization
  starting at LO, i.e., $\alpha_s^0$, up to a power-suppressed error.
  Fixed-order calculations of the hard scattering are appropriate.
\item On the right-hand side, the integral of $W$ is zero.  So the
  integral of the right-hand side is the integral of $Y$ plus the
  error term.
\item But $Y$ is obtained from collinear factorization starting at
  NLO, i.e., $\alpha_s^1$.
\end{itemize}
If we used a fixed order of collinear factorization, there is a
mismatch of orders of $\alpha_s$, and this results in complete mismatch of
sizes even at the highest $Q$.  Recall that the elementary derivation
of the $W+Y$ formula concerned the errors at intermediate $\Tsc{q}$
and did not concern itself explicitly with the integral over
$\Tsc{q}$. 

As regards the integral over all $\Tsc{q}$, we deduce that either we
need resummation of $Y$ to handle the problem, or the error term is
intrinsically very large, or both.  The knowledge that nevertheless
fixed-order calculations with collinear factorization are completely
appropriate for the integrated cross section motivates part of our
method for improving the $W+Y$ method.

%============================================
\section{Our new proposal}

We modify $W$ in two ways, as exhibited in our formula for a modified
$W$ term:
\begin{equation}
\label{eq:W.new}
  W_{\rm New} =
  \Xi\xleft(\frac{\Tsc{q}}{Q}\right)
  \int \frac{\diff[2]{\T{b}}}{(2 \pi)^2}
    e^{i\T{q}\cdot \T{b} } \tilde{W}(b_c(\Tsc{b}),Q).
\end{equation}

First, to avoid problems in $\int \diff[2]{\T{q}} W_{\rm New}$, we
provide the $\tilde{W}$ function with a smooth cutoff at small
$\Tsc{b}$:
\begin{equation}
  b_c(\Tsc{b}) = \sqrt{ \Tsc{b}^2 + \mbox{const}/Q^2 }.
\end{equation}
The integral of \eqref{eq:W.new} over $\Tsc{q}$ is now given by
$\tilde{W}(\Tsc{b})$ at $\Tsc{b}$ of order $1/Q$.  This is correctly
predicted by fixed-order collinear factorization, and agrees with
collinear factorization for the integrated cross section at leading
order.  Higher-order terms bring in the integral of our
correspondingly modified $Y$ term, with well-behaved collinear
expansions.  This prescription is close to that of
\citet{Bozzi:2005wk}.  Their prescription was formulated purely in
terms of resummation calculations in massless QCD.  Our solution
applies to full TMD factorization.  The function $\tilde{W}(b,Q)$ has
the same functional form as before, and involves exactly the same TMD
pdfs and evolution equations; the modification consists in changing
the value used for the transverse position argument in the Fourier
transform, from $\Tsc{b}$ to $b_c(\Tsc{b})$.  At low $\Tsc{q}$, larger
values of $\Tsc{b}$ than $1/Q$ dominate, and then the cutoff at small
$\Tsc{b}$ is unimportant; thus the validity of TMD factorization in
its target region of $\Tsc{q} \ll Q$ is unaffected.

The second change is to impose an explicit cutoff at large $\Tsc{q}$
cutoff, by a factor
\begin{equation}
   \Xi\xleft(\frac{\Tsc{q}}{Q}\right) 
   = \exp \left[ -\left( \frac{q_T}{ \mbox{const.} Q} \right)^{\rm{const.}}
           \right].
\end{equation}
This keeps the modified $W$ term from being significantly nonzero at
such large $\Tsc{q}$ that TMD factorization is totally inapplicable.

Correspondingly, to implement a correct matching, we modify $Y$ to
\begin{equation}
  Y_{\rm New} = X(\Tsc{q}) \times
     \mbox{collinear approx.\ to} 
      \left( \frac{ \diff{\sigma} }{ \diff[4]{q} } - W_{\rm New} \right)
\end{equation}
The second factor is the basic implementation of matching of TMD and
collinear factorization.  But we impose an extra $\Tsc{q}$ cutoff
factor, for example
\begin{equation}
  X(\Tsc{q}/\lambda) = 1 - \exp \left\{ -(\Tsc{q} / \mbox{const.})^{\mbox{const.}} \right\}.
\end{equation}

The reason for the extra cutoffs on $W$ at large $\Tsc{q}$ and on $Y$
at small $\Tsc{q}$ is found in the derivation of the errors in Sec.\
\ref{sec.error}.  That error calculation is only valid when $\Lambda \lesssim
\Tsc{q} \lesssim Q$.  Below that range, we should use only the TMD
factorization term $W$, i.e., $Y$ should then be close to zero.  Above
that range, we should only use collinear factorization, so $W$ should
then be close to zero.

%============================================
\section{Conclusions}

We modified the $W+Y$ formalism, so that
\begin{itemize}
\item The error in $W+Y$ is suppressed by a power of $Q$ for all
  $\Tsc{q}$. 
\item The integral over all $\Tsc{q}$ is now properly behaved with
  respect to collinear factorization for the integrated cross
  section. 
\end{itemize}

Further improvements are undoubtedly possible.  We have tried to
formulate some relevant issues.
Generally, to do better, one needs to ``look under hood'', to ask
questions like:
\begin{itemize}
\item What is the nature of the approximations giving factorization
  (TMD and collinear)?
\item How much do they fail, with proper account of non-perturbative
   properties?
\end{itemize}

These issues are important for subject of this conference, i.e., spin
physics.  This is because we often want to use TMD factorization at
moderate $Q$, notably in the measurement of transverse-spin-dependent
TMD pdfs and fragmentation functions.

One important question for SIDIS, is to determine the appropriate
criteria for what is large and small transverse momentum relative to
$Q$.  Is the appropriate variable $\Tsc{q}$ or $\Tsc{P}$?  I.e., is it
transverse momentum of the virtual photon relative to the detected
hadrons, or is it the transverse-momentum of the detected final-state
hadron relative to the $\gamma^*$ and target?  These two variables differ by
a factor of the fragmentation variable $z$.  One can also ask whether
that was even the right question.

It would also be useful to have the $Y$-term for SIDIS at NNLO, i.e.,
$O(\alpha_s^2)$.  This could be obtained from the results for collinear
factorization for SIDIS at the same order, as reported by Daleo et
al.\ \cite{Daleo:2004pn}, but we are not aware that this has been done
yet.

%============================================

\begin{acknowledgments}
  This work was supported by DOE contracts No.\ DE-AC05-06OR23177
  (A.P., T.R., N.S., B.W.), 
  under which Jefferson Science Associates, LLC operates Jefferson
  Lab,
  No.\ DE-FG02-07ER41460 (L.G.), and No.\ DE-SC0013699 (J.C.), 
  and by the National Science Foundation under Contract No. PHY-1623454 (A.P.).
\end{acknowledgments}

%============================================
\bibliography{jcc}

\end{document}